\documentclass[acmsmall]{acmart}

\usepackage{soul}
\usepackage{hyperref}
\usepackage{multirow}
\usepackage{listings}
\usepackage{fancybox}
\usepackage{enumitem}
\usepackage{graphicx}
\usepackage{color}
\usepackage{array}
\usepackage{amsmath}
\usepackage{xspace}
\usepackage{url}
\usepackage{tikz}
\usepackage{caption}
\usepackage{pgfplots}
\usepackage{balance}
\usepackage{subcaption}
\pgfplotsset{compat=1.18}
\usepgfplotslibrary{groupplots}
\usepackage{pgf-pie}
\usepackage{algorithm}
\usepackage{algorithmic}
\usepackage{adjustbox}

\usepackage{makecell}
\usepackage{tabularx}
\usepackage{tcolorbox}
\makeatletter
\newcommand{\mybox}[1]{%
	\setbox0=\hbox{#1}%
	\setlength{\@tempdima}{\dimexpr\wd0+13pt}%
	\begin{tcolorbox}[boxrule=0.5pt, colback=white, arc=4pt,
		left=6pt,right=6pt,top=6pt,bottom=6pt,boxsep=0pt]
		#1
	\end{tcolorbox}
}

\definecolor{songcolor}{RGB}{191,191,191}

\usepackage[framemethod=tikz]{mdframed}
\definecolor{lightblue}{RGB}{227,242,253}

\newmdenv[
  backgroundcolor=lightblue,
  linecolor=black,
  skipabove=10pt,
  skipbelow=10pt,
  linewidth=1pt,
  innertopmargin=6pt,
  innerbottommargin=6pt,
  innerleftmargin=6pt,
  innerrightmargin=6pt,
  frametitlebackgroundcolor=lightblue,
  frametitlefont=\bfseries,
]{findingbox}

\AtBeginDocument{%
  }

\setcopyright{acmlicensed}
\copyrightyear{2024}
\acmYear{2024}
\acmDOI{XXXXXXX.XXXXXXX}

\begin{document}

\title{Program Slicing in the Era of Large Language Models}

\author{Kimya Khakzad Shahandashti}
\email{kimya@yorku.ca}
\affiliation{%
  \institution{York University}
  \city{Toronto}
  \state{Ontario}
  \country{Canada}
}

\author{Mohammad Mahdi Mohajer}
\email{mmm98@yorku.ca}
\affiliation{%
  \institution{York University}
  \city{Toronto}
  \state{Ontario}
  \country{Canada}
}

\author{Alvine Boaye Belle}
\email{alvine.belle@lassonde.yorku.ca}
\affiliation{%
  \institution{York University}
  \city{Toronto}
  \state{Ontario}
  \country{Canada}
}

\author{Song Wang}
\email{wangsong@yorku.ca}
\affiliation{%
  \institution{York University}
  \city{Toronto}
  \state{Ontario}
  \country{Canada}
}

\author{Hadi Hemmati}
\email{hemmati@yorku.ca}
\affiliation{%
  \institution{York University}
  \city{Toronto}
  \state{Ontario}
  \country{Canada}
}

\renewcommand{\shortauthors}{Khakzad et al.}

\begin{abstract}
Program slicing is a critical technique in software engineering, enabling developers to isolate relevant portions of code for tasks such as bug detection, code comprehension, and debugging. In this study, we investigate the application of large language models (LLMs) to both static and dynamic program slicing, with a focus on Java programs. We evaluate the performance of four state-of-the-art LLMs—GPT-4o, GPT-3.5 Turbo, Llama-2, and Gemma-7B— by leveraging advanced prompting techniques, including few-shot learning and chain-of-thought reasoning. 
Using a dataset of 100 Java programs derived from LeetCode problems, our experiments reveal that GPT-4o performs the best in both static and dynamic slicing across other LLMs, achieving an accuracy of 60.84\% and 59.69\%, respectively. Our results also show that the LLMs we experimented with are yet to achieve reasonable performance for either static slicing or dynamic slicing. Through a rigorous manual analysis, we developed a taxonomy of root causes and failure locations to explore the unsuccessful cases in more depth. We identified \textit{Complex Control Flow} as the most frequent root cause of failures, with the majority of issues occurring in \textit{Variable Declarations and Assignments} locations. To improve the performance of LLMs, we further examined two independent strategies for prompting guided by our taxonomy, including \textit{prompt crafting}, which involved refining the prompts to better guide the LLM through the slicing process and \textit{iterative prompting}, where the model receives feedback on the root cause and location of the failure and re-generates its responses. Our evaluation shows these two prompting enhancement approaches can improve accuracy by 4\% and 3.9\%, respectively. 
\end{abstract}




\maketitle

\section{Introduction}

Program slicing~\cite{weiser1984program,sasirekha2011program} is a crucial technique in software engineering, enabling early bug detection, enhancing code comprehension, and improving debugging effectiveness~\cite{tip1995survey}. By isolating relevant code portions based on a \textit{slicing criterion}, such as a variable or line number~\cite{weiser1984program}, developers can better understand complex systems and pinpoint issues. This process is instrumental in simplifying the maintenance and evolution of complex systems~\cite{ishio2003program}. In general, two primary approaches to program slicing exist, i.e., static slicing \cite{acharya2011practical}, which analyzes code without execution, and dynamic slicing \cite{korel1988dynamic}, which considers program execution to capture runtime behavior~\cite{ agrawal1990dynamic}. Traditional program slicing tools~\cite{galindo2022program, mohapatra2014slicing, zhang2021sympas, slicer4j} have been extensively studied and continually refined, making them reliable for a wide range of slicing tasks. However, the emergence of large language models (LLMs) in natural language processing (NLP) has introduced new possibilities in software engineering~\cite{hou2023large, ahmed2023few, sun2023automatic, chen2021evaluating, fan2023large}, particularly for tasks like bug detection~\cite{alrashedy2023language} and static analysis~\cite{mohajer2024effectiveness}. Despite growing interest in LLMs, there has yet to be a comprehensive evaluation of their effectiveness in performing program slicing. This study aims to bridge that gap by assessing the performance of LLMs in program slicing tasks. In this work, we explore the capabilities of LLMs in both static and dynamic program slicing, with a focus on Java programs. The novelty of this study lies in the development of a taxonomy of unsuccessful slicing cases, providing a structured framework for understanding where and why LLMs fail in generating accurate slices. This taxonomy represents a critical step toward improving LLM-based slicing techniques and serves as a foundation for future research in the field. We also propose strategies, guided by this taxonomy, to improve the accuracy of LLMs in program slicing tasks. Our study is driven by the following research questions (RQs):

\vspace{5pt}
\noindent \textbf{RQ1: How do LLMs perform in static program slicing?}

\noindent \textbf{RQ2: How do LLMs perform in dynamic program slicing?}

\noindent \textbf{RQ3: What are the characteristics of unsuccessful static program slices?}

\noindent \textbf{RQ4: How do prompt enhancement strategies improve the accuracy of LLM-based static program slicing?}
\vspace{5pt}

To address these research questions, we utilize a dataset derived from solutions to LeetCode problems~\cite{hartford2024leetcode}, which offers various programs across multiple programming languages. For this study, we focus on a 
randomly selected subset of 100 Java programs, preprocessed for compatibility with traditional static and dynamic slicing tools. We employ advanced prompting techniques, such as one-shot learning~\cite{brown2020language} and chain-of-thought reasoning~\cite{wei2022chain}, using four state-of-the-art models: GPT-4o~\cite{openai2024gpt4o}, GPT-3.5 Turbo~\cite{openai2024gpt35turbo}, Llama-2-7B-Chat~\cite{meta2024llama2chat}, and Gemma-7B~\cite{google2024gemma7b}. These models are evaluated against traditional slicing tools to assess their accuracy in generating program slices. Our experiments reveal that GPT-4o achieved the highest performance in both static and dynamic slicing tasks, with an accuracy of 60.84\% in static slicing and 59.69\% in dynamic slicing. However, our results also indicate that the subject LLMs have yet to reach reasonable performance levels for either task, underscoring the inherent challenges LLMs face in generating precise program slices.

Furthermore, we conduct an in-depth manual analysis of the unsuccessful slices to explore why and where LLMs fail to perform slicing, categorizing the root causes and identifying failure locations within the code. This analysis culminated in the creation of a novel taxonomy of failures in LLM-based program slicing, which constitutes a key contribution of this study. We concentrated our taxonomy-building efforts on static slicing, primarily for its predictability
and the more well-defined dependencies it involves.
The taxonomy categorizes root causes into three key subcategories: \textit{Lack of Logic Understanding}, which includes Conditional Statements, Loops, and Method Invocations; \textit{Code Complexity}, encompassing \textit{Complex Control Flow} and \textit{Ambiguity in Code}; and \textit{Model-specific Constraints}. Among these, \textit{Complex Control Flow} emerged as the most frequent root cause of failures. Additionally, we identified six categories regarding the locations of failures: \textit{Conditional Statements}, \textit{Loop Constructs}, \textit{Method Invocations and Returns}, \textit{Variable Declarations and Assignments}, \textit{Class Declarations}, and \textit{Imports}. \textit{Variable Declarations and Assignments} were identified as the most common failure location, where incorrect or incomplete slices were generated. These detailed categories provide valuable insights into where and why LLMs fall short in program slicing, serving as a foundation for future improvements and targeted strategies. Leveraging these insights, we examined two independent prompt enhancement strategies to improve accuracy for static slicing. The first strategy, \textit{prompt crafting}, involves refining the structure and content of the prompts based on our taxonomy of failures.
This approach helps guide the LLMs more effectively through the slicing process and results in an accuracy improvement of 4\% across GPT-4o. The second strategy, \textit{iterative prompting}, provides the model with feedback on the root cause and location of its errors and allows it to re-generate its response. This technique yielded a 
3.9\% improvement in accuracy. 
\newline

Our contributions are four-fold: 
\begin{itemize} 

\item \textbf{Evaluation of LLM-based program slicing}: We evaluate the performance of state-of-the-art LLMs in static and dynamic slicing tasks. 

\item \textbf{A taxonomy of unsuccessful program slicing}: We provide a novel taxonomy of the root causes and common fault locations behind unsuccessful static program slices.

\item \textbf{Prompt strategies guided by our taxonomy to enhance LLM-based static program slicing}: We {further examined} enhanced prompt crafting and iterative prompting strategies, guided by the taxonomy, to improve the accuracy of LLMs in static program slicing.

\item \textbf{Artifacts.} We release the dataset and source code of our experiments to help other researchers replicate and extend our study\footnote{\url{https://anonymous.4open.science/r/ProgramSlicingLLMs-4E60}}.

\end{itemize}

\section{Background and Related Work}

\subsection{Program Slicing}
Program slicing, introduced by Weiser \cite{weiser1984program}, is a fundamental technique in software engineering that isolates parts of a program influencing a particular computation, referred to as the \textit{slicing criterion} \cite{binkley1996program}. This method is widely employed in debugging, testing, and maintenance tasks, allowing developers to hone in on the relevant portions of code while excluding unrelated parts \cite{ahmed2021mandoline}. There are two primary approaches to program slicing: static and dynamic. Static slicing analyzes the program without executing it, providing insights into the overall code structure and dependencies. This approach is particularly useful for optimization and enhancing code comprehension \cite{tip1995survey}. In contrast, dynamic slicing takes into account the program's execution with specific input values, making it highly effective for debugging and addressing runtime issues \cite{korel1988dynamic, agrawal1990dynamic}. Although dynamic slicing offers more precision in capturing execution-specific behavior, it is resource-intensive due to the need for actual execution data and test cases \cite{korel1997dynamic}. Both slicing techniques have been extensively studied and refined over the years. Static slicing tends to be more efficient but may produce larger-than-needed slices since it includes all potential execution paths, whereas dynamic slicing is more precise but requires considerable resources for execution \cite{tip1995survey, korel1988dynamic}.

When generating a program slice, two key types can be constructed: backward slicing and forward slicing \cite{harman2001overview}. A backward slice captures all the statements that can potentially affect the slicing criterion, making it valuable for understanding the origins of values or bugs. In contrast, a forward slice identifies the statements that are affected by the slicing criterion, which is useful for tracking the downstream impacts of specific computations. This study focuses exclusively on backward slicing, as it is particularly effective for debugging by tracing the dependencies leading up to a specified behavior. 


\subsection{LLMs in SE Tasks}
Recently, significant research has explored the potential of LLMs in Software Engineering (SE) tasks. Fan et al. \cite{fan2023large} provide a comprehensive survey of LLMs for SE tasks. Several studies have focused on utilizing LLMs for software testing and fuzzing \cite{deng2023large, hu2023augmenting, xie2023chatunitest,yuan2023no}. For instance, Deng et al. \cite{deng2023large} introduce TitanFuzz, a novel approach that harnesses LLMs to generate and mutate input programs for fuzzing deep learning libraries, achieving significantly higher code coverage and identifying numerous previously unknown bugs compared to traditional fuzzing techniques. Additionally, some research has concentrated on using LLMs in Requirements Engineering tasks \cite{zhang2023evaluation, luitel2024improving, khakzad2024assessing, shahandashti2024using}. Zhang et al. \cite{zhang2023evaluation} evaluate ChatGPT's performance on requirements analysis tasks to gain insights into the impact of LLMs on both research and the practice of natural language processing within requirements engineering. In the realm of program analysis, several studies have examined the application of LLMs \cite{mohajer2024effectiveness, chapman2024interleaving,fang2023largellm, li2023hitchhiker}. For example, Mohajer et al. \cite{mohajer2024effectiveness} assess ChatGPT's effectiveness in static analysis tasks, such as static bug detection and false positive warning removal. Yadavally et al. \cite{yadavally2024learning} introduce NS-Slicer which is a notable study that leverages pre-trained language models, specifically targeting the task of static program slicing. NS-Slicer uses GraphCodeBERT \cite{guo2020graphcodebert} and CodeBERT \cite{feng2020codebert} to predict static slices for both complete and partial code. This study serves as an important reference point for our research. We plan to use NS-Slicer as a baseline for evaluating the effectiveness of our LLM-based approach.

\section{Study Design}

This section outlines the methodology and setup used to evaluate LLMs for program slicing tasks. Our study involves generating program slices using LLMs, comparing them against ground-truth slices produced by traditional slicing tools, and assessing their accuracy using various metrics. 
In addition to the automated evaluation, we have conducted a manual analysis of unsuccessful slices, which helped us identify common root causes and failure locations. As a result of our analysis, we developed a novel taxonomy of unsuccessful slicing cases regarding the root causes and locations. Based on the taxonomy, we further examined two prompting strategies to improve the performance of LLMs, including enhanced prompt crafting and iterative prompting techniques. These strategies were aimed at addressing the specific failure patterns identified during the manual analysis and resulted in improvements in the models' accuracy in program slicing tasks.

\begin{figure}[t]
\centering
\includegraphics[width=0.85\textwidth]{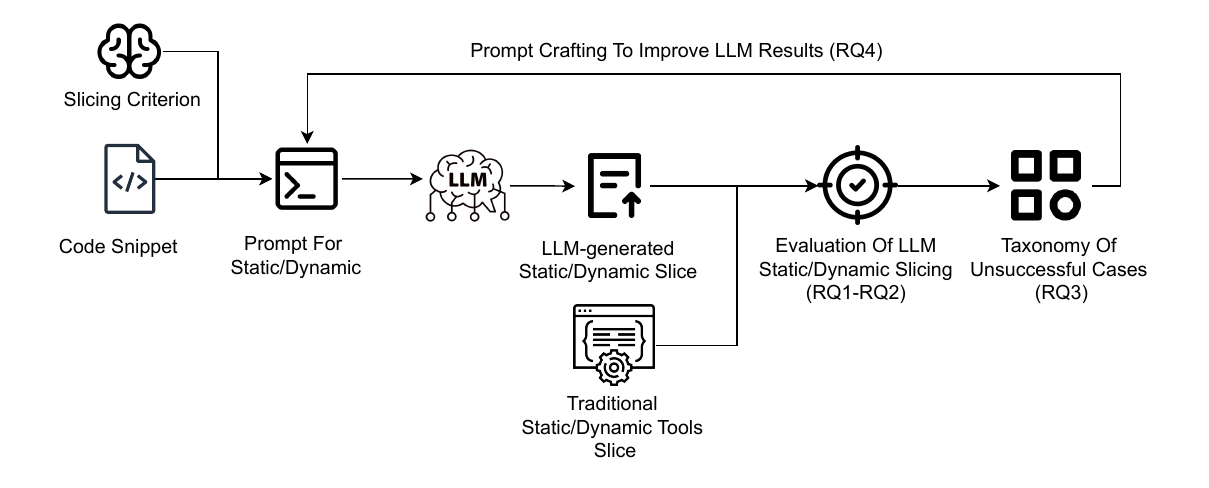}
\vspace{-0.1in}
\caption{Overview of this study}
\label{fig:overview}
\end{figure}

\subsection{Overview}

Figure \ref{fig:overview} illustrates the overall process of our LLM-based program slicing study. The process begins with two key inputs: the \textit{Slicing Criterion} (which specifies the variable or line to slice) and a code snippet. Depending on whether static or dynamic slicing is required, a corresponding prompt is generated and fed into an LLM. The LLM then produces a program slice based on the prompt. In parallel, to accelerate the process of labeling ground truth, we leverage traditional program slicing tools (i.e., we use JavaSlicer~\cite{javaslicer} for static slicing and Slicer4J~\cite{slicer4j} for dynamic slicing) to generate slices that serve as the candidate ground truth for each program. 
We then manually verified the slices generated by these traditional tools, cross-referencing them with program logic to create the ground truth slicing for each program. 
Next, we evaluate the LLM-generated slices for both static and dynamic slicing by comparing them with the manually verified ground-truth slices (\textbf{RQ1} and \textbf{RQ2}). For cases where the LLM-generated slices are unsuccessful, we perform a detailed analysis to identify and categorize the underlying root causes and failure locations (\textbf{RQ3}). Finally, we apply iterative prompt refinement, adjusting and improving prompts based on the identified failures to enhance the LLM's slicing performance (\textbf{RQ4}). This iterative process allows us to address the challenges and limitations identified in previous steps, ultimately improving the accuracy and robustness of LLM-generated program slices.

\subsection{Dataset Description}
The dataset used in this study consists of a diverse collection of solutions and explanations for various LeetCode problems, covering multiple programming languages such as Python, Java, C, and JavaScript. For our experiments, we focused exclusively on Java, randomly selecting 100 samples. 
This selection was made to balance the computational cost of running LLMs and the scope of the analysis while ensuring a representative sample size. The dataset was preprocessed to be compatible with both static and dynamic analysis tools. For static slicing, we added necessary libraries, adjusted function names, and ensured the code was properly structured for slicing tools. For dynamic slicing, we manually inserted a \texttt{main} function into each program to enable execution-based analysis. 
Moreover, our \textit{slicing criterion} focused on manually selecting the key variables and lines of code that were central to the program's logic. 

\subsection{Labeling Ground Truth}
To evaluate the performance of LLMs, it was essential to establish ground-truth program slices for comparison. We adopted a two-step process to generate and validate these ground-truth slices. First, we used traditional program slicing tools to automatically generate candidate ground-truth slices. These tools provide accurate slices based on well-established slicing algorithms, serving as a benchmark for assessing the LLM-generated slices. In the second step, we manually verified and, when necessary, adapted these slices to ensure correctness and alignment with the program's intended behavior. 
We employed two primary tools: JavaSlicer and Slicer4J. JavaSlicer~\cite{javaslicer} is a static program slicing tool that analyzes Java bytecode using the system dependence graph (SDG) to extract program slices. 
Slicer4J~\cite{slicer4j}, on the other hand, is a dynamic program slicing tool designed specifically for Java programs. It captures runtime information to produce program slices reflecting the program's execution. 


\subsection{Experimental LLMs}
In this study, we selected four state-of-the-art large language models (LLMs) for evaluation: Llama-2-7B-Chat, Gemma-7B, GPT-3.5 Turbo, and GPT-4o. \textbf{Llama-2-7B-Chat}, developed by Meta AI~\cite{touvron2023llama}, is fine-tuned for chat-based applications and coding assistance, containing 7 billion parameters. It is designed to provide high-quality conversational and coding capabilities, making it particularly useful for tasks requiring the generation of programming instructions. \textbf{Gemma-7B}, developed by Google as part of the Gemma family~\cite{team2024gemma}, is an open-source, decoder-only text-to-text model optimized for English language tasks, with a context length of up to 8192 tokens. Its versatility allows it to excel in tasks such as summarization, question answering, and reasoning. \textbf{GPT-3.5 Turbo}, a variant of the GPT-3.5 series from OpenAI~\cite{openai}, is designed for high efficiency and strong performance in text generation tasks. It is well-suited for real-world applications that require high-quality text output at a lower computational cost. Lastly, \textbf{GPT-4o}, the flagship model in the GPT-4 series by OpenAI, is engineered to handle complex tasks with superior accuracy. This model is known for its powerful capabilities in both natural language and programming code generation, making it an ideal choice for our programming-related tasks. Table \ref{tab:llms} summarizes the characteristics of the mentioned LLMs. Each model was used with its default settings for our experiments, with GPT-4o configured at a temperature of 0.7 and Llama-2-7B-Chat at a temperature of 0.8, ensuring consistency across all evaluations.

\begin{table}[t!]
\centering
\caption{Overview of experimental LLMs}
\label{tab:llms}
\begin{tabular}{lcccc}
\toprule
\textbf{Model} & \textbf{Llama-2-7B-Chat} & \textbf{Gemma-7B} & \textbf{GPT-3.5 Turbo} & \textbf{GPT-4o} \\
\midrule
Size & 7B & 7B & - & - \\ 
Context Window & 4K & 8K & 4K & 8K \\ 
Release Date & Jul 2023 & Feb 2024 & March 2023 & March 2024 \\ 
\bottomrule
\end{tabular}
\end{table}

\subsection{Evaluation Metrics}
To capture the models' performance at the slice level, we used two metrics to evaluate our experiments, both adopted from the work of Yadavally et al. \cite{yadavally2024learning}. The first metric, \textbf{Exact-Match Accuracy (Accuracy-EM)}, evaluates the model's performance by counting the number of times the predicted slices exactly match the ground-truth slices. The second metric, \textbf{Dependence Accuracy (Accuracy-D)}, measures how accurately the inter-statement dependencies are predicted. Accuracy-D is calculated as the ratio of correctly predicted dependencies to actual dependencies for all slicing criteria in a given program.

\section{LLM-Based Program Slicing}

\subsection{Prompting Techniques}

In this study, we employed several commonly used prompting techniques to evaluate the performance of the LLMs. First, \textbf{zero-shot prompting} involves providing the model with a prompt without any examples or additional context, relying solely on the model's pre-trained knowledge to generate a response based on the input alone~\cite{kojima2022large}. Next, we used \textbf{one-shot prompting}, where the model is given a prompt along with a single example to guide its response. This technique helps the model understand the task better by showcasing the desired output format, allowing it to leverage its ability to learn from minimal data to improve task performance~\cite{gu2021ppt}. Lastly, we employed \textbf{Chain-of-Thought (CoT)} prompting \cite{wei2022chain}, which breaks down complex tasks into intermediate reasoning steps. This method guides the model through a sequence of logical steps to arrive at the final answer, enhancing its ability to handle tasks that require multi-step reasoning~\cite{wang2022self,zhang2022automatic}.

\subsection{Experimental Setup}
We designed three experiments to investigate the effectiveness of different prompting techniques in generating accurate and relevant program slices. 

\noindent \textbf{Experiment 1: zero-shot.} This experiment evaluates the ability of the model to generate program slices using zero-shot prompting, where the model is not provided with any specific examples or additional context. The model relies solely on its pre-trained knowledge to perform the task.

\noindent \textbf{Experiment 2: one-shot.} This experiment incorporates a one-shot learning approach, where the model is given a single example to guide its response. In this setup, the model is provided with one example of a program slice with its respective output, but no additional context is given. We selected the specific example that was introduced in the traditional program slicing tools. This example is a pair consisting of a \textit{<code\_snippet, slicing\_criterion, output>}. This approach ensures that the example used in the one-shot learning setup is both relevant and representative of the typical slicing tasks the LLM is expected to perform.

\noindent \textbf{Experiment 3: one-shot with CoT.} This experiment combines one-shot learning with the CoT prompting technique. In this setup, the model is provided with a single example that includes a program slice along with a logical sequence of steps or reasoning chains that guide its slicing output. We used the same methodology as in Experiment 2 for selecting the example, relying on the example introduced in the traditional program slicing tools. This example is a pair consisting of a \textit{<code\_snippet, slicing\_criterion, reasoning, output>}. By breaking down the problem into these intermediate steps through the reasoning chain, the Chain-of-Thought technique helps the model to better understand and execute the task \cite{wei2022chain}.

\subsection{Prompting Templates}
This section presents the prompt templates used for RQ1 and RQ2, which guide the LLM in performing slicing tasks while maintaining consistency in responses across different experiments. 
Since we employ various prompting strategies depending on the experiment, the structure of the prompt changes accordingly, particularly regarding the inclusion of examples. For zero-shot, no example is provided within the prompt. In one-shot, we include an example comprising a code snippet, the associated slicing criterion, and the expected output to guide the LLM's response. In one-shot with CoT, the prompt is further expanded to include an example with a code snippet, the slicing criterion, a reasoning section detailing the steps needed to derive the output, and the expected output itself.

\subsubsection{RQ1: Static Slicing Template}

Table \ref{tab:templates_static} presents the prompt template for static slicing. It provides a structured approach, guiding the LLM to identify all lines that impact the Slicing Criterion and then trace backward to earlier lines. The template ensures that the model fully utilizes its capabilities to generate accurate and comprehensive slices by clearly defining the Slicing Criterion format. To maintain consistency and facilitate easy comparison with ground truth data, the output is formatted in JSON, simplifying the parsing process for subsequent analysis.
\begin{table*}[h!]
\centering
\footnotesize
\caption{Prompt template for RQ1}
\label{tab:templates_static}
\begin{tabularx}{\textwidth}{X}
\hline
\textbf{Prompt Template} \\ 
\hline
You are an AI assistant specialized in performing backward static slicing for Java programs.
You are provided with a Java code snippet with the line numbers and a Slicing Criterion. 

\textbf{Important Notes:}

- Your task is to identify all lines in the program that may affect the value of the Slicing Criterion variable, following a backward slicing approach.

- You need to first locate the line of the last instruction that affects the Slicing Criterion Variable in the code (That line is always in the output), then trace backward to find all relevant lines.

\textbf{Slicing Criterion Format:}
\textbf{<Provide Slicing Criterion Format>}

\textbf{Output Format:}

- An array of line numbers in plain JSON without any markdown in the following format: 
\{\{"output": ["line\_number1", "line\_number2"]\}\}

\textbf{<Provide Example with Code Snippet, Slicing Criterion, and Output for One-shot Experiment. Include Reasoning for One-shot with CoT Prompting. Omit Example for Zero-shot.>} 

\textbf{Task:}

Now, based on the provided Java program and Slicing Criterion, generate the output in the specified format. 

\textbf{Slicing Criterion}: 
\textbf{<Provide Slicing Criterion>}

\textbf{Program}: \textbf{<Provide Code Snippet>}

\textbf{Output}: \\

\hline
\end{tabularx}
\end{table*}

\subsubsection{RQ2: Dynamic Slicing Template}
Table \ref{tab:templates_dynamic} presents the prompt template for dynamic slicing. It directs the LLM to focus on the Slicing Criterion, typically corresponding to a critical line such as the return statement in the main function, and to trace backward through the code to identify all lines that directly or indirectly influence it. This template ensures the model accurately captures the runtime dependencies essential for dynamic slicing. We format the output in JSON to enable seamless parsing and direct comparison with ground-truth data during subsequent analysis.

\begin{table*}[h!]
\centering
\footnotesize
\caption{Prompt template for RQ2}
\label{tab:templates_dynamic}
\begin{tabularx}{\textwidth}{X}
\hline\textbf{Prompt Template} \\ 
\hline
You are an AI assistant specialized in performing backward dynamic slicing for Java programs. 
You are provided with a Java code snippet, which includes line numbers, and a Slicing Criterion line number.

\textbf{Important Notes:}

- The Slicing Criterion line number corresponds to the return statement in the \texttt{main} function.

- Your task is to start with the Slicing Criterion line number itself in the output and then trace backward through the code to identify all relevant lines that directly or indirectly influence the value at the Slicing Criterion.

\textbf{Output Format:}

- An array of line numbers in plain JSON without any markdown in the following format: 
\{\{"output": ["line\_number1", "line\_number2"]\}\}

\textbf{<Provide Example with Code Snippet, Slicing Criterion, and Output for One-shot Experiment. Include Reasoning for One-shot with CoT Prompting. Omit Example for Zero-shot.>} 

\textbf{Task:}

Now, based on the provided Java program and Slicing Criterion, generate the output in the specified format.

\textbf{Slicing Criterion}: 
\textbf{<Provide Slicing Criterion>}

\textbf{Program}: \textbf{<Provide Code Snippet>}

\textbf{Output}: \\ 
\hline
\end{tabularx}
\end{table*}

\subsection{RQ1: Effectiveness of LLMs on Static Program Slicing}

Table \ref{tab:static} presents the performance metrics for static program slicing across three experiments (i.e., zero-shot, one-shot, and one-shot with Chain-of-Thought) for selected LLMs, including GPT-4o, GPT-3.5 Turbo, Llama-2, and Gemma-7B. The metrics evaluated are Accuracy-D and Accuracy-EM. Moreover, to mitigate the effects of the inherently non-deterministic nature of LLMs, we ran each experiment three times and used the 
{average} for each model to present its performance. To benchmark our LLM-based approach, we employed NS-Slicer~\cite{yadavally2024learning} as the baseline for static slicing of Java programs. NS-Slicer leverages both CodeBERT and GraphCodeBERT, and their pre-trained models are publicly available. We carefully preprocessed our dataset to ensure compatibility with NS-Slicer for execution. However, due to the complexity of certain programs, we were unable to run 15\% of the dataset. Despite this, NS-Slicer effectively predicts static program slices, making it a valuable benchmark for comparison with our LLM-based approach. Table \ref{tab:nsslicer} presents the performance of NS-Slicer on our dataset. It is worth noting that the performance of NS-Slicer in our study differs from the results reported in its original paper. This discrepancy can be attributed to several factors. First, our dataset is different, as we selected Java programs from LeetCode, which vary in complexity compared to the dataset used in the original NS-Slicer study, which is IBM’s Project CodeNet dataset \cite{puri2021codenet}. Another possible factor is the preprocessing steps we performed to ensure compatibility with the traditional tools, which may have introduced variations in the code format that could influence the performance.
\begin{table*}[t!]
\centering
\caption{Effectiveness of LLMs on static program slicing across different prompt strategies} 
\vspace{-0.1in}
\label{tab:static}
\begin{tabular}{llcccccc}
\toprule
\multirow{2}{*}{\textbf{Model}} & \multicolumn{2}{c}{\textbf{Zero-shot}} & \multicolumn{2}{c}{\textbf{One-shot}} & \multicolumn{2}{c}{\textbf{One-shot with CoT}} \\
\cmidrule(lr){2-3} \cmidrule(lr){4-5} \cmidrule(lr){6-7} 
 & \textbf{Acc-D} & \textbf{Acc-EM} & \textbf{Acc-D} & \textbf{Acc-EM} & \textbf{Acc-D} & \textbf{Acc-EM} \\
\midrule
GPT-4o        & 30.91\% & 0\%  & 54.09\% & 4.0\%  & \textbf{60.84\%} & \textbf{7.33\%}  \\
GPT-3.5 Turbo & 29.60\% & 0\%  & 37.33\% & 0\%  & 46.43\% & 1\%  \\
Llama-2       & 4.75\%  & 0\%  & 10.54\% & 0\%  & 17.57\% & 0\%  \\
Gemma-7B      & 38.65\% & 0\%  & 35.22\% & 0\%  & 43.43\% & 0.66\%  \\
\bottomrule
\end{tabular}
\end{table*}

\begin{table}[t!]
\caption{Effectiveness of NS-Slicer on our dataset}\vspace{-0.1in}
\label{tab:nsslicer}
    \begin{tabular}{lcc}
    \toprule
    \textbf{Model} & \textbf{Acc-D} & \textbf{Acc-EM} \\
    \midrule
    NS-Slicer (CodeBERT)      & 60.52\% & 0\% \\
    NS-Slicer (GraphCodeBERT) & 55.33\% & 0\% \\
    \bottomrule
    \end{tabular}
\end{table}

As we can see from Table \ref{tab:static}, GPT-4o demonstrated the best performance across all models, particularly in the one-shot with CoT prompting strategy, achieving an Accuracy-D of \textbf{60.84\%} and an Accuracy-EM of \textbf{7.33\%}. This is notable because GPT-4o performs better than NS-Slicer concerning exact match accuracy, as neither NS-Slicer (CodeBERT) nor NS-Slicer (GraphCodeBERT) generated any exact matches. Regarding Accuracy-D, GPT-4o also slightly outperforms NS-Slicer (CodeBERT) and NS-Slicer (GraphCodeBERT). NS-Slicer (GraphCodeBERT) shows lower performance, achieving an Accuracy-D of 55.33\%. GPT-3.5 Turbo also showed improvement across the experiments, with its highest performance in the one-shot with CoT prompting strategy, achieving an Accuracy-D of 46.43\%. Llama-2 consistently performed weaker than the other models, with its best result in the one-shot with CoT strategy reaching only 17.57\% for Accuracy-D. Gemma-7B exhibited moderate performance, performing best in the one-shot with CoT prompting strategy, achieving an Accuracy-D of 43.43\%.

\begin{findingbox}
        \textbf{Key Findings:} GPT-4o emerged as the best-performing model for static slicing, especially under the one-shot with chain-of-thought prompting strategy, which resulted in the Accuracy-D of 60.84\% and an Accuracy-EM of 7.33\%. NS-Slicer (CodeBERT) achieved comparable slice Accuracy-D of 60.52\% but could not generate any exact matches. 
\end{findingbox}

\subsection{RQ2: Effectiveness of LLMs on Dynamic Program Slicing}

Table \ref{tab:dynamic} presents the performance metrics for dynamic program slicing across different prompt strategies (i.e., zero-shot, one-shot, and one-shot with Chain-of-Thought) for various LLMs, including GPT-4o, GPT-3.5 Turbo, Llama-2, and Gemma-7B, in Java. The evaluated metrics are Accuracy-D and Accuracy-EM. Similar to RQ1, we ran each experiment three times and selected the 
average for each model to report its performance. The findings from our analysis of dynamic slicing results indicate that GPT-4o consistently outperforms the other models in dynamic slicing, achieving the highest Accuracy-D score of \textbf{59.69\%} under the zero-shot prompting strategy. However, GPT-4o, like the other models, struggles to achieve exact matches, as indicated by its Accuracy-EM of 0.0\% across all experiments. GPT-3.5 Turbo also performs relatively well, particularly in the zero-shot strategy, where it achieves an Accuracy-D of 44.84\%, though its performance drops slightly in one-shot and one-shot with chain-of-thought strategies. Llama-2 performs the best under the one-shot with chain-of-thought strategy with an Accuracy-D of 36.72\%, although it shows lower performance in other strategies. Gemma-7B demonstrates mixed performance, achieving its best Accuracy-D of 41.90\% in the one-shot strategy, but its performance drops significantly in the one-shot with chain-of-thought strategy, reaching 25.49\%. The performance inconsistency across experiments can be attributed to the nature of dynamic slicing tasks. Dynamic slicing relies heavily on execution contexts and runtime behavior, which may not be adequately captured in all prompting strategies. The additional context provided in the prompts appears to introduce confusion. This suggests that, for dynamic slicing, more context does not always lead to better performance, likely due to the complexity of runtime behavior and the challenges faced by models when handling additional information in such cases.

\begin{table*}[t!]
\caption{Effectiveness of different prompting techniques for dynamic program slicing}
\vspace{-0.1in}
\label{tab:dynamic}
\begin{tabular}{llcccccc}
\toprule
\multirow{2}{*}{\textbf{Model}} & \multicolumn{2}{c}{\textbf{Zero-shot}} & \multicolumn{2}{c}{\textbf{One-shot}} & \multicolumn{2}{c}{\textbf{One-shot with CoT}} \\
\cmidrule(lr){2-3} \cmidrule(lr){4-5} \cmidrule(lr){6-7}
 & \textbf{Acc-D} & \textbf{Acc-EM} & \textbf{Acc-D} & \textbf{Acc-EM} & \textbf{Acc-D} & \textbf{Acc-EM} \\
\midrule
GPT-4o & \textbf{59.69\%} & 0\% & 41.77\% & 0\% & 58.62\% & 0\% \\
GPT-3.5 Turbo & 44.84\% & 0\% & 42.44\% & 0\% & 36.16\% & 0\% \\
Llama-2 & 25.26\% & 0\% & 26.85\% & 0\% & 36.72\% & 0\% \\
Gemma-7B & 33.41\% & 0\% & 41.90\% & 0\% & 25.49\% & 0\% \\
\bottomrule
\end{tabular}
\end{table*}

\begin{findingbox}
        \textbf{Key Findings:} GPT-4o remains the most effective model for dynamic slicing, particularly under the zero-shot strategy, achieving an Accuracy-D of \textbf{59.69\%}. However, all models across all experiments struggle with exact matches, as indicated by the consistent 0\% Accuracy-EM.  
\end{findingbox}

\subsection{Comparison of Static and Dynamic Slicing}


In this section, we compare the effectiveness of static and dynamic slicing across the evaluated LLMs. To assess the statistical significance of the performance difference between static and dynamic slicing, we applied the Mann-Whitney U test \cite{macfarland2016mann}, a non-parametric statistical test that does not assume normal distribution. This test is particularly suited for our data, given the small sample size and the fact that accuracy values may not follow a normal distribution. We used the Accuracy-D from Tables \ref{tab:static} and \ref{tab:dynamic}, which represent the performance of LLMs on static and dynamic slicing, respectively. 
The results of the Mann-Whitney U test yielded a U statistic of \textbf{63.0} and a p-value of \textbf{0.62}. As the p-value exceeds the commonly accepted threshold of 0.05, we concluded that there is no statistically significant difference between the effectiveness of static slicing and dynamic slicing. However, it is important to note that while no significant difference was found in Accuracy-D, all models demonstrated an advantage in static slicing compared to dynamic slicing when looking at Accuracy-EM. 

\section{RQ3: Taxonomy of Unsuccessful Program Slicing}

In RQ1 and RQ2, our experimental results show that most of the subject LLMs are yet to achieve a reasonable performance for either static slicing or dynamic slicing. In this section, we set out to demystify the reasons behind these unsuccessful slicing cases. We concentrated our taxonomy-building efforts on static slicing, primarily for its predictability and the more well-defined dependencies it involves. Moreover, the complexities of dynamic slicing—such as the need for runtime execution—introduce additional challenges that make manual analysis and taxonomy building more difficult. The runtime behavior adds layers of variability that complicate the identification of consistent failure patterns. Focusing first on static slicing, we establish a robust taxonomy that addresses core issues, which can later be adapted and extended to more complex, dynamic slicing scenarios. Our study identified GPT-4o as the best-performing model in static slicing, producing slices with higher Accuracy-D compared to other models. To streamline the manual effort of labeling unsuccessful cases for taxonomy creation, we focused on GPT-4o under the prompting strategy of one-shot with CoT, which consistently generated higher-quality slices. Additionally, we concentrated on the best iteration of the model's output, allowing us to manage the manual effort more effectively while still targeting the most promising results. By concentrating on the best-performing model and its best iterations, we ensured that the taxonomy reflects the most nuanced and sophisticated challenges encountered in program slicing, rather than simply cataloging basic errors from less capable models.

\subsection{Methodology}
To systematically identify unsuccessful slicing, we focused on the cases where the Accuracy-D of the slicing produced by the LLM was not 100\%. This approach allowed us to ensure a comprehensive analysis of the LLM’s performance, capturing every instance where the model did not perfectly replicate the ground-truth slices. By including all non-perfect slices, we could thoroughly investigate the specific areas where the LLMs struggled, ensuring that no potential issue was overlooked. A total of \textbf{92} unsuccessful cases required manual analysis. Two of the authors, each with over 5 years of experience in software development, independently reviewed the unsuccessful slicing cases to ensure objective classification. The authors carefully analyzed each case, focusing on identifying the root causes and error locations within the LLM-generated slices. On average, each sample took approximately 5 minutes to assess, which involved understanding the code, comparing the LLM-generated slice with the ground truth, and labeling both the root cause and error location. When a new root cause or fault location category emerged that was not accounted for in the taxonomy, the authors temporarily paused their labeling efforts and held meetings with each other. These discussions allowed for the validation of new categories, updates to the taxonomy, and re-labeling of affected cases, ensuring consistency and accuracy throughout the process. After completing their reviews, the two experts met to discuss and resolve any discrepancies in their classifications. Through collaborative discussion, they reached a consensus on each case. This rigorous process resulted in a refined taxonomy that categorizes root causes into three main subcategories: Lack of Logic Understanding (including Conditional Statements, Loops, and Method Invocations), Code Complexity (including Complex Control Flow and Ambiguity in Code), and Model-specific Constraints. In addition, the authors identified six distinct failure location categories: Conditional Statements, Loop Constructs, Method Invocations and Returns, Variable Declarations and Assignments, and Class Declarations and Imports.

\subsection{Taxonomy of Fault Locations}
\begin{figure}[t!]
\centering
\includegraphics[width=\textwidth]{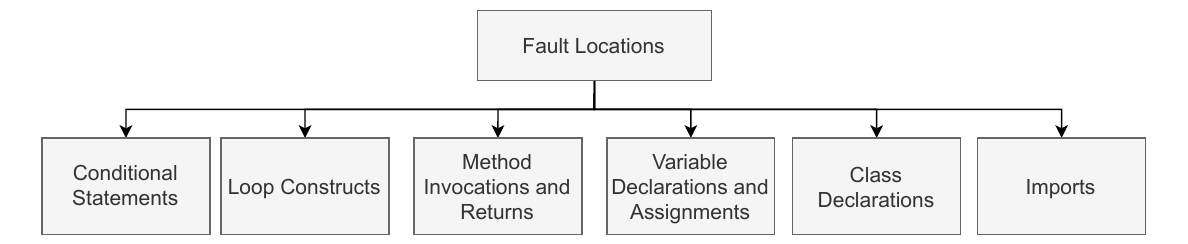}
\vspace{-0.1in}
\caption{Categories of locations where slicing faults occur}
\label{fig:location}
\end{figure}

Faults within a codebase can occur at various critical locations, each presenting unique challenges for accurate program slicing by LLMs. Identifying and categorizing these fault locations is essential for understanding the slicing failures and developing strategies to improve the accuracy of LLM-generated slices. Figure \ref{fig:location} illustrates the six categories of fault locations in unsuccessful slicing cases. Below, we describe the primary categories of fault locations that commonly affect slicing accuracy. We also provide an example to illustrate each category. 

\noindent \textbf{A1. Conditional Statements.} Conditional statements, such as \texttt{if}, \texttt{else}, and \texttt{switch}, are common locations for slicing faults. These faults typically arise when the LLM incorrectly evaluates the branching logic, leading to the misinterpretation of which branches are relevant to the slicing criterion. As a result, the generated slice may either over-include irrelevant branches or omit necessary ones, causing incomplete or inaccurate slices.


\noindent \textbf{A2. Loop Constructs.} Loops, including \texttt{for}, \texttt{while}, and \texttt{do-while} loops, are common locations for slicing errors. These faults lead to slices that either exclude essential loop iterations or unnecessarily include irrelevant ones, failing to correctly propagate the slicing criterion across the loop structure.


\noindent \textbf{A3. Method Invocations and Returns.} Method Invocations and return statements are another critical location for faults. 


\noindent \textbf{A4. Variable Declarations and Assignments.} Variable declarations and assignments are frequent locations of slicing failures. Ambiguities in how variables are declared or initialized can lead to data dependency errors, where the LLM fail to accurately track variable states throughout the program.


\noindent \textbf{A5. Class Declarations.} Faults occurring in class declarations can have an impact on slicing accuracy.


\noindent \textbf{A6. Imports.} Imports represent a location where slicing errors occur. When external modules or libraries are imported, the LLM may not fully understand how these imports impact the program, leading to incomplete slices that fail to include the necessary functionality brought in by the imports.


\subsection{Taxonomy of Root Causes}
\begin{figure}[t!]
\centering
\includegraphics[width=\textwidth]{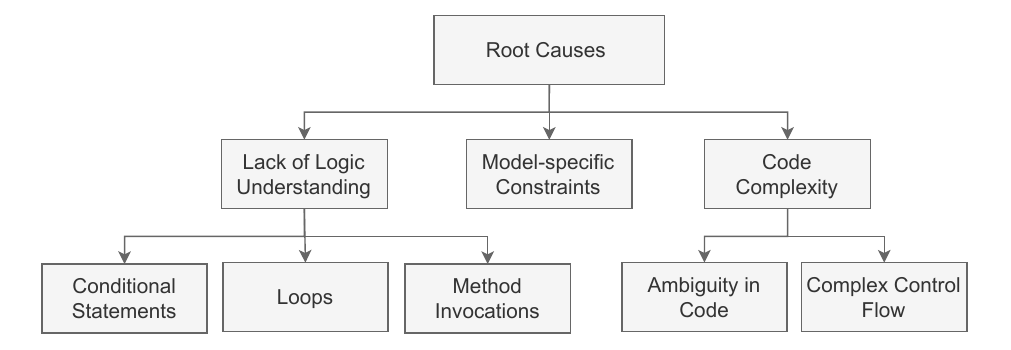}
\vspace{-0.1in}
\caption{Categories of the root causes}
\label{fig:taxonomy}
\end{figure}

Figure \ref{fig:taxonomy} illustrates the categories of root causes for unsuccessful cases, which consists of three subcategories, i.e., Lack of Logic Understanding (Section~\ref{sec:5.2.1}), Code Complexity (Section~\ref{sec:5.2.2}), and Model-specific Constraints (Section~\ref{sec:5.2.3}). 

\subsubsection{Lack of Logic Understanding} 
\label{sec:5.2.1}

The analysis revealed that unsuccessful slicing cases often arise from the LLM's inability to fully understand the logical structure of the program. These logic-related failures typically occur in code segments that involve decision-making and control flow. The LLM struggles to accurately reason about the dependencies in these regions, which hinders the proper propagation of slicing criteria and leads to incomplete or incorrect slices. The key logical challenges where LLMs frequently struggle include conditional statements, loops, and method invocations.


\noindent \textbf{B1. Conditional Statements.} Errors related to conditional statements, such as \texttt{if} and \texttt{switch}, were commonly observed. These issues typically arise when the LLM fails to capture the entire control flow relevant to the slicing criterion. For instance, the LLM may incorrectly include only the \texttt{if} branch, while omitting necessary \texttt{else} or \texttt{else if} branches, resulting in an incomplete slice. In this example, when slicing based on the \texttt{free} variable, both branches influence its value. However, the LLM might incorrectly slice only the \texttt{if} branch, leading to an incomplete representation of the program logic.

\begin{lstlisting}[language=Java]
if (target[i] > free) {
    req += target[i] - free;
    free = target[i];
} else if (target[i] < free) {
    free = target[i];
}
// Expected: The slice should include both branches when slicing on the 'free' variable.
// LLM Error: The slice may only include the 'if' branch and ignore the 'else' branch.
\end{lstlisting}

\noindent \textbf{B2. Loops.} Slicing errors in loops (such as \texttt{while} and \texttt{for}) frequently arise from the LLM's difficulty in understanding complex loop conditions, invariants, or exit criteria. This often results in slices that are either incomplete or incorrectly exclude significant parts of the loop’s execution, such as intermediate calculations or modifications to data structures. In this example, both the \texttt{sum} and \texttt{queue} statements are crucial for the program’s logic. However, when slicing based on the \texttt{queue} variable, the LLM might exclude the \texttt{sum} operation, leading to an incomplete slice and missing the interdependencies within the loop.

\begin{lstlisting}[language=Java]
for (int num : nums) {
    sum = sum.add(BigDecimal.valueOf(num).divide(BigDecimal.valueOf(2)));
    queue.add(BigDecimal.valueOf(-num));
}
// Expected: The slice should include both statements in the loop when slicing on the 'queue' variable.
// LLM Error: The slice may only include the 'queue' statement, ignoring the 'sum' operation.
\end{lstlisting}

\noindent \textbf{B3. Method Invocations.} Method invocations, especially those involving built-in data structures, often present significant challenges for LLMs. The LLM may fail to propagate the slicing criterion through these method calls, which can affect the internal state of data structures. These methods often manipulate key variables, and failing to track their effects can result in incomplete slices.  In this case, the \texttt{add()}, \texttt{remove()}, and \texttt{size()} method invocations directly influence the internal state of the \texttt{numbers} list. When slicing based on the \texttt{size} variable, the LLM needs to capture all these method invocations. Failure to do so would result in an incomplete representation of the program, as key changes to the \texttt{numbers} list would be missed, affecting the final output.

\begin{lstlisting}[language=Java]
List<Integer> numbers = new ArrayList<>();
numbers.add(5);
numbers.remove(0);
int size = numbers.size();
return size;
// Expected: The slice should include the 'add', 'remove', and 'size' method invocations as they modify or access the internal state of 'numbers'.
// LLM Error: The slice may fail to capture one or more method invocations, missing their impact on the 'numbers' list.
\end{lstlisting}

\subsubsection{Code Complexity}
\label{sec:5.2.2}
The analysis also identified several root causes related to the complexity of the code that contributed to unsuccessful slicing. These root causes were consistently linked to the LLM’s limitations in understanding and processing complex programming constructs. Two main categories of root causes were identified:

\noindent \textbf{C1. Ambiguity in Code.} Ambiguous code constructs were a significant source of errors in program slicing. These ambiguities often arise when the code can be interpreted in multiple ways, causing the LLM to make incorrect assumptions about how the slicing criterion should propagate. A common scenario involves variables that are initialized across multiple lines, or expressions that involve complex or unclear logic, leading to incomplete or incorrect slices. In this example, the value of `result[i]` is determined by a multi-line ternary expression, which introduces ambiguity. The slicing criterion focuses on the variable `result[i]`, but due to the way the expression spans several lines, the LLM might fail to capture the entire initialization process. 

\begin{lstlisting}[language=Java]
result[i] = 
s.length() > (intLength + 1) / 2 
    ? -1
    : someFunction(s, intLength);
// Expected: The slice should correctly propagate the value of 'result[i]'.
// LLM Error: The LLM might fail to capture the full expression due to the multi-line structure.
\end{lstlisting}

\noindent \textbf{C2. Complex Control Flow.} The LLM exhibited difficulties when handling complex control flows, particularly in the case of nested loops and conditionals, where dependencies and data flow become intricate and challenging to track. Although loops and conditionals are already part of the sub-category of Lack of Logic Understanding, we introduce this category because the complexity increases significantly when these constructs are nested.
This added complexity often led to incomplete or incorrect slices, as the LLM failed to accurately propagate the slicing criterion across all relevant paths within the code.  The nested structure of the loops, as shown above, can confuse the LLM, particularly when tracking how the variables `z`, `i`, and `y` depend on each other.

\begin{lstlisting}[language=Java]
for (int z = m - 1; z >= 0; --z) {
    for (int i = 0; i <= z; ++i) {
        int y = z - i;
    }
}
// Expected: The slice should capture the relationship between 'z', 'i', and 'y' across all iterations.
// LLM Error: The LLM fail to track dependencies between the variables, especially in nested loops.
\end{lstlisting}

\subsubsection{Model-Specific Constraints}
\label{sec:5.2.3}
Several unsuccessful slices stemmed from inherent limitations in the design of the LLMs used. These model-specific constraints often resulted in slicing errors and reduced the overall effectiveness of the LLM in accurately processing complex code. The LLM's internal context window can cause errors in cases where the code being sliced exceeds the model’s token limit, leading to incomplete slices as critical code dependencies are truncated. Additionally, when text is intermixed with code, such as in comments, documentation, or string literals, the LLM can mistakenly include irrelevant text as part of the slice, resulting in incorrect slices
. Parsing failures also occur due to the LLM's reliance on JSON-based outputs for processing results. In some cases, the JSON parser can misinterpret or fail to process the LLM's output correctly, especially when the model generates malformed or incomplete JSON structures. These model-specific constraints represent inherent challenges that limit the LLM’s ability to slice complex programs fully and accurately.


\subsection{Discussion}
\label{sec:5.2.4}

\begin{figure}[h!]
\centering
\includegraphics[width=\textwidth]{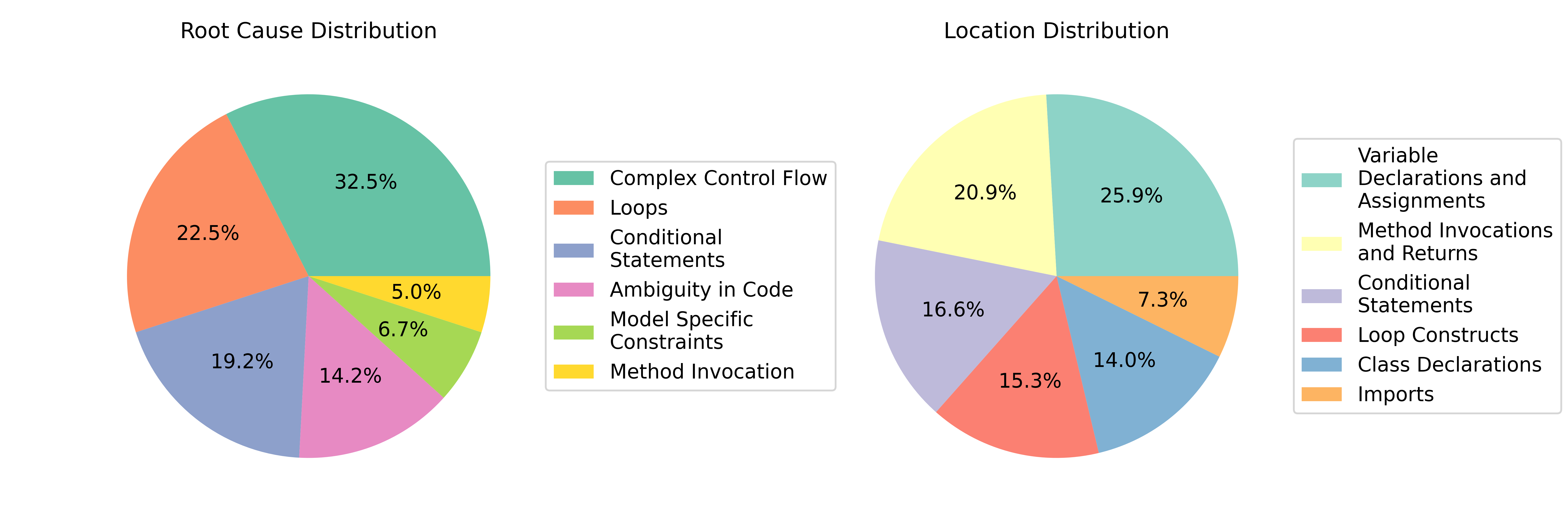}
\caption{Distribution of unsuccessful slicing cases} 
\label{fig:distribution}
\end{figure}
\begin{figure}[t!]
\centering
\includegraphics[width=0.6\textwidth]{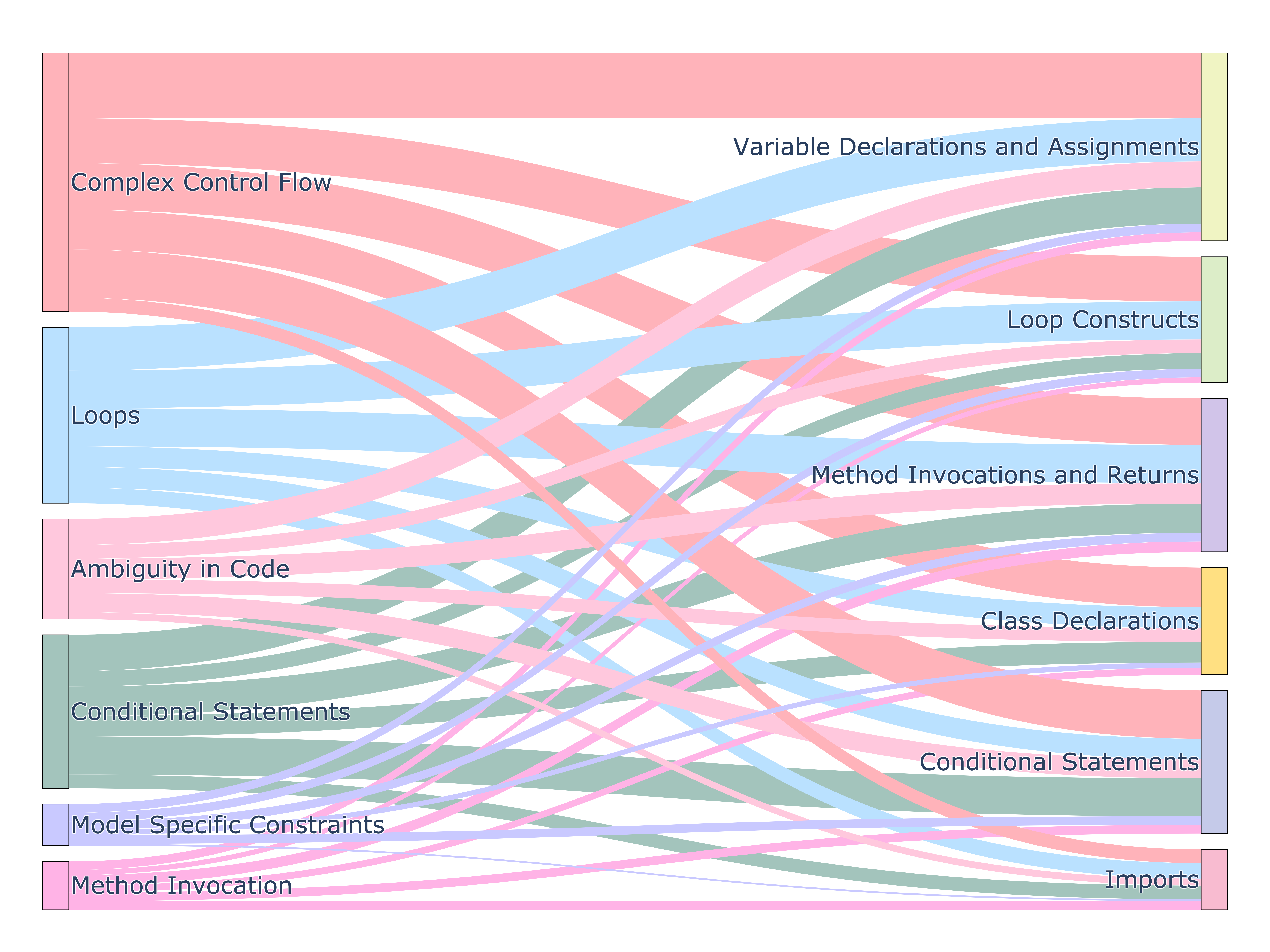}
\caption{Mapping between root causes and failure locations} 
\label{fig:sankey}
\end{figure}

Figure \ref{fig:distribution} illustrates the distribution of root causes and location across these slices, highlighting the potential impact each category has on the accuracy of LLM-generated slices. Out of the 92 unsuccessful slicing cases that required manual analysis, several key patterns emerged. The most frequent \textbf{root cause} of errors is \textit{Complex Control Flow}, identified in 39 cases. This suggests that LLMs struggle with handling intricate control flow structures such as nested loops and conditionals, which are prevalent in more complex codebases. In terms of \textbf{location}, the most common location of errors is \textit{Variable Declarations and Assignments}, identified in 78 cases. This points to significant challenges LLMs face in correctly slicing code around variable declaration and assignments. These findings highlight key areas where LLMs struggle, particularly when dealing with complex control structures and dependencies. Understanding these challenges is crucial for refining LLM-based slicing techniques and improving their overall accuracy. We also created a Sankey diagram to illustrate the relationship between root causes and failure locations in LLM-based program slicing, as shown in Figure \ref{fig:sankey}. The diagram provides a depiction of how root causes are linked to fault locations. For example, the diagram highlights that the majority of errors related to \textit{Complex Control Flow} happen in \textit{Loop Constructs} and \textit{Variable Declarations and Assignments} locations.

\begin{findingbox}
    \textbf{Key Findings:} Through manual analysis, we proposed a taxonomy for root causes and locations of failures in LLM-based static program slicing. Our findings reveal that \textit{Complex Control Flow} is the most common root cause, while \textit{Variable Declarations and Assignments} is the most frequent error location.
\end{findingbox}

\section{RQ4: Strategies for Improvement}

Rangeet et al.~\cite{pan2024lost} proposed prompt crafting and iterative prompting to improve LLM-based code translation tasks. Motivated by~\cite{pan2024lost}, in this section, we set out to examine the effectiveness of these two approaches on LLM-based static program slicing. In this context, we refer to the results from RQ1 in one-shot with CoT prompting as \textit{Vanilla}, which served as the baseline for comparison as it has the best Accuracy-D among other experiments. Our goal is to build upon this baseline and enhance performance through targeted improvements. This taxonomy enables researchers to accurately guide LLMs to avoid the most common root causes and fault locations leading to unsuccessful slicing across four subject LLMs: GPT-4, GPT-3.5 Turbo, Llama-2, and Gemma-7B. 

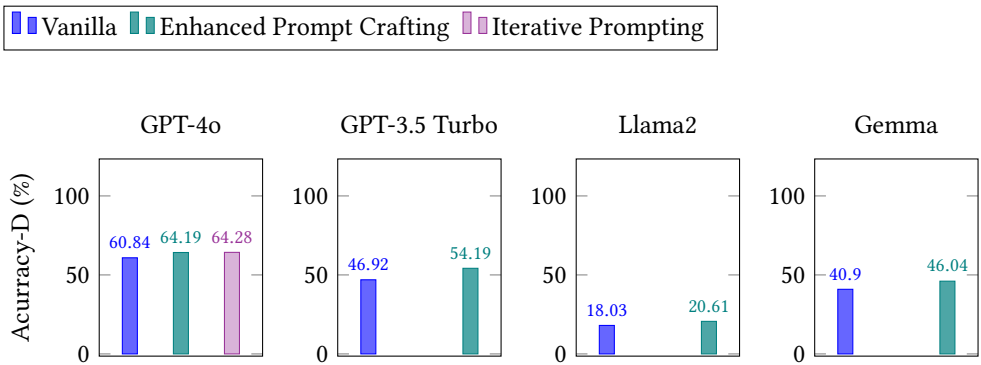
\begin{figure}[t]
\pgfplotsset{compat=1.18}
\centering
\begin{tikzpicture}
\begin{groupplot}[
    group style={
        group name=my plots, 
        group size=4 by 1, 
        xlabels at=edge bottom, 
        ylabels at=edge left,
        vertical sep = 0.5cm
    },
    height=0.3\linewidth, 
    width=0.27\linewidth, 
    ylabel={Acurracy-D (\%)},
    ybar, 
    enlargelimits=0.30, 
    symbolic x coords={A, B, C}, 
    xtick=\empty,
]

\nextgroupplot[
    title={GPT-4o},
    bar width = 0.2cm,
    nodes near coords,
    nodes near coords style={
        font=\fontsize{7pt}{6pt}\selectfont 
    },
    legend style={
        at={(1.60,1.55)}, 
        anchor=south, 
        legend columns=3, 
        /tikz/every even column/.append style={column sep=0.1cm},
        legend cell align={left}
    },
    ymin=22,
    ymax=100
]
\addplot+[bar shift=0cm, fill=blue!60, draw=blue!100, text=blue!100] coordinates {(A,60.84)}; \addlegendentry{Vanilla}
\addplot+[bar shift=0cm, fill=teal!70, draw=teal!100, text=teal!100] coordinates {(B,64.19)}; \addlegendentry{Enhanced Prompt Crafting}
\addplot+[bar shift=0cm, fill=violet!30, draw=violet!80, text=violet!80] coordinates {(C,64.28)}; \addlegendentry{Iterative Prompting}

\nextgroupplot[
    title={GPT-3.5 Turbo},
    bar width = 0.2cm,
    nodes near coords,
    nodes near coords style={
        font=\fontsize{7pt}{6pt}\selectfont 
    },
    legend style={
        at={(1.60,1.55)}, 
        anchor=south, 
        legend columns=3, 
        /tikz/every even column/.append style={column sep=0.1cm},
        legend cell align={left}
    },
    ymin=22,
    ymax=100
]
\addplot+[bar shift=0cm, fill=blue!60, draw=blue!100, text=blue!100] coordinates {(A,46.92)}; 
\addplot+[bar shift=0cm, fill=teal!70, draw=teal!100, text=teal!100] coordinates {(B,54.19)}; 

\nextgroupplot[
    title={Llama2},
    bar width = 0.2cm,
    nodes near coords,
    nodes near coords style={
        font=\fontsize{7pt}{6pt}\selectfont 
    },
    legend style={
        at={(1.60,1.55)}, 
        anchor=south, 
        legend columns=3, 
        /tikz/every even column/.append style={column sep=0.1cm},
        legend cell align={left}
    },
    ymin=22,
    ymax=100
]
\addplot+[bar shift=0cm, fill=blue!60, draw=blue!100, text=blue!100] coordinates {(A,18.03)}; 
\addplot+[bar shift=0cm, fill=teal!70, draw=teal!100, text=teal!100] coordinates {(B,20.61)}; 

\nextgroupplot[
    title={Gemma},
    bar width = 0.2cm,
    nodes near coords,
    nodes near coords style={
        font=\fontsize{7pt}{6pt}\selectfont 
    },
    legend style={
        at={(1.60,1.55)}, 
        anchor=south, 
        legend columns=3, 
        /tikz/every even column/.append style={column sep=0.1cm},
        legend cell align={left}
    },
    ymin=22,
    ymax=100
]
\addplot+[bar shift=0cm, fill=blue!60, draw=blue!100, text=blue!100] coordinates {(A,40.90)}; 
\addplot+[bar shift=0cm, fill=teal!70, draw=teal!100, text=teal!100] coordinates {(B,46.04)};

\end{groupplot}

\end{tikzpicture}
\caption{Comparison of accuracy improvements for different LLMs (GPT-4o, GPT-3.5 Turbo, Llama2, and Gemma) under various strategies: Vanilla, Enhanced Prompt Crafting, and Iterative Prompting}
\label{fig:improvements}
\end{figure}
\noindent \textbf{1. Enhanced Prompt Crafting:} 
Refining the prompts used to guide LLMs has been shown to enhance their accuracy \cite{white2023prompt}. As highlighted in Section \ref{sec:5.2.4}, the most common root cause for failures is \textit{Complex Control Flow}, while the most frequent error location occurs in \textit{Variable Declarations and Assignments}. To mitigate these issues, we enriched the prompts by adding an example that includes a code snippet, detailed reasoning steps explaining how the slicing should be correctly performed, and the expected output. This example contains common failure scenarios identified in our manual analysis, such as \textit{Complex Control Flow} and \textit{Variable Declaration and Assignment}, where we identified LLMs as having historically struggled. The results demonstrate that this enhanced prompt crafting approach consistently improves performance across different LLMs. For GPT-4o, accuracy increased by approximately 4\%, while GPT-3.5 Turbo showed a similar improvement, with accuracy rising by 6.4\%. Gemma-7B also experienced a gain in performance, with an increase of 5.1\%, and Llama2 saw an improvement of 2.5\%. These results indicate that prompt crafting, when guided by insights from our taxonomy of common failure points, can significantly enhance the LLMs' ability to handle complex static program slicing tasks. Figure \ref{fig:improvements} demonstrates the effectiveness of this enhanced prompt crafting strategy, showcasing the Accuracy-D gains achieved across the selected LLMs compared to the Vanilla approach.

\noindent \textbf{2. Iterative Prompting:} 
Building on the approach outlined in \cite{stechly2023gpt}, we employed iterative prompting to improve the performance of LLM-generated program slices. This technique was applied exclusively to GPT-4o after we identified specific failure cases through manual analysis. In this process, we provided feedback directly in the prompt, explaining why the initial slices were incorrect by detailing the root causes and failure locations identified during the manual review. By explicitly guiding the LLM to address these issues, we instructed it to avoid the same pitfalls while re-generating the slices. This iterative feedback loop allowed the model to focus on the areas where it had previously struggled, enabling it to refine its performance in a targeted manner. In this single iteration, we observed an increase in Accuracy-D by \textbf{3.9\%}, demonstrating the effectiveness of targeted feedback in addressing the specific issues the model encountered. While only one iteration was performed, further iterations could improve the model's performance even more. Each iteration would involve identifying new failed cases, labeling them, and providing feedback based on root causes and failure locations, ultimately improving the LLM’s ability to generate accurate slices. Although the iterative approach is independent of enhanced prompt crafting, both techniques complement each other by targeting different aspects of the LLM’s limitations.
It is important to note that, while this approach can improve performance, it may not be practical as it requires a human-in-the-loop to verify the results of each iteration to provide feedback. Figure \ref{fig:improvements} demonstrates the effectiveness of iterative prompting for GPT-4o, highlighting the gains achieved compared to the Vanilla approach.

\begin{findingbox}
    \textbf{Key Findings:} Our investigation into enhancement strategies revealed that both \textit{enhanced prompt crafting} and \textit{iterative prompting}, guided by our proposed taxonomy, improved the Accuracy-D of LLM-based static program slicing. Enhanced prompt crafting increased the Accuracy-D by up to 4\% for GPT-4o, while iterative prompting improved GPT-4o's Accuracy-D by 3.9\%.
\end{findingbox}

\section{Threats to Validity}

In this section, we outline potential threats to the validity of our study and explain the steps we took to address them.

\noindent \textbf{External Validity}. A key threat to external validity is the limited range of programming languages we used in our experiments. Our study focused exclusively on Java, which may restrict the generalizability of our findings to other languages. One reason for this choice is the availability of traditional slicing tools like JavaSlicer and Slicer4J, which are well-tested and maintained. Expanding the study to other languages, such as C, would require identifying or developing reliable slicing tools for those languages. Another external validity concern relates to the limited number of LLMs selected for our experiments. Future work could involve a broader range of LLMs to determine if our findings hold across different architectures and model versions. Another potential threat arises from using the same dataset to develop the taxonomy and evaluate the approach. This overlap could affect the generalizability of our findings. To mitigate this issue, future work could involve using separate datasets: one for developing the taxonomy and another for evaluating the approach.

\noindent \textbf{Internal Validity}. One internal threat arises from the possibility that the LLMs had seen our dataset during their training. To mitigate this, we preprocessed the dataset to ensure that the LLMs had not encountered it before. This step helped prevent the models from generating responses based on previously seen data, ensuring a more accurate evaluation of their slicing capabilities on unseen code. Furthermore, our results could be impacted by potential errors in our automation scripts. To minimize this risk, we performed extensive testing of the scripts and conducted spot checks to verify the correctness of the outputs. Additionally, we have made our artifacts publicly accessible to support the review and replication of our findings \cite{ProgramSlicingLLMsReplicationPackage}.

\noindent \textbf{Construct Validity}. A threat to construct validity stems from the manual analysis of unsuccessful slices. The subjective nature of manual labeling can introduce bias. To address this, we employed two independent reviewers to evaluate the results, followed by a discussion to resolve any discrepancies. This collaborative process ensured a more consistent and objective analysis. Another potential issue lies in the reliability of the traditional program slicing tools we used as the ground truth. Any inaccuracies in these tools could affect our evaluation of the LLM-generated slices. We mitigated this by using well-established tools such as JavaSlicer and Slicer4J, which are trusted for their accuracy and reliability.
\section{Conclusion}
Program slicing is essential in software engineering, aiding in tasks like bug detection. In this study, we set out to investigate how LLMs could be applied to program slicing, specifically focusing on both static and dynamic slicing for Java programs. Our findings reveal that GPT-4o demonstrated the best performance among the models tested, but challenges remain, particularly in achieving exact matches in more complex scenarios. Recognizing the need to understand these failures better, we examined the unsuccessful cases and characterized the root causes and locations of such failures. Our taxonomies offer a structured framework for researchers to identify common patterns in unsuccessful static program slices, providing insights into where LLMs struggle most in handling program-slicing tasks. Furthermore, we leveraged this taxonomy to further examine two prompt enhancement strategies, i.e., enhanced prompt crafting and iterative prompting.

For future work, we plan to expand the scope of this study to include other programming languages to provide a more comprehensive evaluation of LLM performance. Additionally, we plan to explore alternative prompt engineering techniques to improve model accuracy.

\section{Data Availability}
All the scripts and data to reproduce our experiments can be found in our replication package \cite{ProgramSlicingLLMsReplicationPackage}.  




\bibliographystyle{ACM-Reference-Format}
\bibliography{main}

\end{document}